\documentstyle[prl,aps,epsfig,multicol]{revtex}
\begin{document}
\def\be{\begin{equation}}
\def\ee{\end{equation}}
\def\bearr{\begin{eqnarray}}
\def\eearr{\end{eqnarray}}

\draft

\title{Revisiting the Hanbury Brown-Twiss Setup for Phase Fluctuating Bose Gases}

\author
{Tarun Kanti Ghosh}

\address
{The Abdus Salam International Centre for Theoretical
Physics, Strada Costiera 11, 34014 Trieste, Italy.}

\date{\today}
\maketitle

\begin{abstract}
The Hanbury Brown-Twiss experiment has proved to be an effective means of measuring two-point
correlation function of identical particles.
We analyze experimental observation of stripes formation 
of a phase fluctuating Bose-Einstein condensates in a highly elongated 3D traps 
[Dettmer {\em et al.}, Phys. Rev. Lett. {\bf 87}, 160406 (2001)]
by means of axial two-point correlation functions. 
We also predict that the stripes are present in quasi-1D Bose gas in the mean-field as
well as in the hard-core bosons regimes.
These stripes can be realized by measuring the axial two-point correlation functions by using the Bragg 
interferometric method which is similar to the original Hanbury Brown and Twiss experimental setup. 
\end{abstract}


\begin{multicols}{2}[]
\section{Introduction}
Since the pioneering works on the realization of
Bose-Einstein condensates (BEC) \cite{rmp} of alkali-atoms,
a great variety of experimental and theoretical investigations have
probed the macroscopic phase coherence of confined quantum gases.
For a trapped 3D BEC well below the transition temperature $T_c$, experiments have confirmed the
macroscopic phase coherence by measuring the correlation length which is equal to the condensate size
\cite{coh}. However, the phase coherence 
strongly depends on the shape of the confining potential which can be control at will.
It was theoretically proposed that the axial phase fluctuations of an elongated 3D BEC can be very large
in the equilibrium state, where the density fluctuations are strongly suppressed \cite{petrov}.
The axial phase coherence length in the elongated systems can be smaller than the axial
size and this is referred as the quasi-condensates.
Similarly, quasi-1D Bose gas in the mean-field regime \cite{gorlitz} forms a quasi-condensate \cite{pet1} with a 
large phase fluctuations even at temperatures as low as $0.1 T_c $ \cite{tkg}.
The quasi-1D Bose gas behaves like a hard-core bosons (commonly known as Tonks-Girardeau Gas) \cite{lieb} when the 
strength of the
two-body potential is very strong which has been observed experimentally \cite{bloch}. The hard-core bosonic systems
have a very strong vacuum phase fluctuations which prevents from being a (quasi-)condensate even at $T=0 $ and
mimicking the exclusion principle for fermions.
The most striking features of the phase fluctuating 3D BEC is
that the stripes formation in the axial density profile of the system \cite{exp}. 

The celebrated Hanbury Brown and Twiss (HBT) experiment \cite{hbt} in which the spatial second-order
correlation function $C_2 (z) $ of a light source is characterized by measuring the
correlations of intensity fluctuations in the wave field. The original idea of the HBT experiment
is that by measuring the intensity correlations between two separated beams, they essentially 
compared the intensities at two different points in the unseperated beam \cite{baym}. 
The equal-times second-order correlation function provides information on the relative spatial 
distribution of pairs of identical particles.
Therefore, the stripes formation can be understand by analyzing the second-order correlation
function if it shows oscillatory behavior. 

In this Letter, we understand the experimental observations 
of the stripes formation in the axial density profile of a phase fluctuating 3D BEC by means of 
the axial two-particle correlation functions.
Next, we propose that the same stripes formation in the axial density profile
can be observed in quasi-1D Bose gas in the mean-field as well as in the
hard-core bosons regimes. 
We also discuss how to realize the stripes formation by a suitable choices
of the Bragg pulse in the Bragg interferometer which is similar to the HBT
experiment.

{\em Two-point correlation function:}
We consider a 3D BEC confined in an elongated harmonic trap, where the repulsive mean-field 
interaction energy exceeds the radial $(\hbar \omega_0)$ and the axial $(\hbar \omega_z)$ trap 
energies. At $T=0 $, the density profile has the well-known form 
$ n_0(\rho,z) = (\mu /g) (1-\rho^2/R_0^2-z^2/Z_0^2)$, where 
$ \mu = 0.5 \hbar \omega_z [15 Na \omega_0^2/a_z \omega_z^2]^{2/5}$ is 
the zero-temperature chemical potential, 
and $ g = 4\pi a \hbar^2/m $ is the two-body interaction strength. Also, $ R_0 = \sqrt{2\mu/m \omega_0^2}$ 
and $Z_0 = \sqrt{2\mu/m \omega_z^2}$ are the radial and the axial size of the condensate, respectively.
Due to the repulsive mean-field energy, density fluctuations are strongly suppressed in a
trapped BEC. Therefore, the bosonic field operator describing the condensate can be written
in the form $ \hat \psi ({\bf r}) = \sqrt{n_0({\bf r})} exp[i \hat \phi ({\bf r})] $,
where the phase operator $ \hat \phi ({\bf r}) $ is defined as
\be \label{phase}
\hat \phi ({\bf r},t) = \sum_{\nu} \sqrt{\frac{2g}{\hbar \omega_{\nu}}} \psi_{\nu}( {\bf r}) e^{-i\omega_{\nu} t} 
\hat 
\alpha_{\nu} + H.c. 
\ee
Here, $  \hat \alpha_{\nu} $ is the annihilation operator of the quasiparticle excitation with quantum numbers
$ \nu $ and energy $ \hbar \omega_{\nu} $; $ \psi_{\nu} $ is the corresponding quasiparticles normalized wave
functions.

The normalized two-particle correlation function
is defined as
\bearr
C_2[\{{\bf r}_i \}] & = & \prod_{i=1}^{4} [\sqrt{n_0({\bf r}_i)}]^{-1} < \hat \psi^{\dag} ({\bf r}_1) \hat
\psi^{\dag}
({\bf r}_2) \hat \psi ({\bf r}_3) \hat \psi ({\bf r}_4) > \nonumber  \\
& = & e^{-\frac{1}{2} <[ \hat \phi ({\bf r}_1) + \hat \phi ({\bf r}_2) - \hat \phi ({\bf r}_3) - \hat \phi ({\bf
r}_4)]^2>}
= e^{-\frac{1}{2} F(\{ {\bf r}_i \})}.
\eearr
Due to the strong suppression of the density fluctuations, the normalized density correlation function
of the trapped condensate is constant, {\em i.e.} $ C_2(r_1,r_2,r_2,r_1) =1 $. Therefore,
a simple measurement of the density correlations  in the condensate is not enough to describe
coherence properties. However, by measuring the density correlations in the interference pattern generated by two 
spatially displaced copies of a parent BEC, it is possible to correlate the bosonic field operator 
$ \hat \psi ({\bf r}) $ at four different positions and  extract the $ C_2(r_1,r_2,r_3,r_4) $ 
\cite{cacci}.

{\em 3D cigar-shaped BEC}:
As described in Ref. \cite{exp},
the 3D condensate consists of $ N = 5 \times 10^5$ atoms of $^{87}$Rb.
The radial and axial trapping frequencies are $\omega_0 =2 \pi \times 365$ Hz
and $\omega_z =2 \pi \times 14$ Hz ($a_z = \sqrt{\hbar/m \omega_z} = 2.879 \mu m$), respectively.
The condensate is already elongated along the longitudinal $z$ axis due to the 
aspect  ratio $ \lambda = \omega_0/\omega_z = 26$.
In addition, further radial compression or expansion of the condensate are obtained by applying
the superimposed blue detuned optical dipole trap to the magnetic trap. In the experiment \cite{exp}, they 
performed the
measurements for the radial trapping frequencies $\omega_0 $ between $ 2 \pi \times 138$ Hz and
$ 2 \pi \times 715$ Hz corresponding to aspect ratios $ \lambda $ between 9.8 and 51, respectively.

The excitations of a cigar-shaped BEC can be divided into two regimes:
``low-energy" axial excitations with energy $ \hbar \omega_z \leq E_a << \hbar \omega_0$,
and ``high-energy" radial excitations with $ E_r \geq \hbar \omega_0 >> \hbar \omega_z$.
As pointed out in Ref. \cite{petrov} that 
the low-energy axial excitations have wave-lengths larger than $R_0$ and exhibit a pronounced 
1D behavior which gives the most important contribution to the low-energy axial phase fluctuations.
The low-energy axial modes
have the energy spectrum $ \epsilon_j = \hbar \omega_z \sqrt{j(j+3)/4}$ \cite{strin}. 
The normalized wave functions  $\psi_j $ of these quasiparticle modes have the form 
\be 
\psi_j({\bf r}) = \sqrt{\frac{(j+2)(2j+3)}{8\pi(j+1) R_0^2 Z_0 }} 
P_j^{(1,1)}(\tilde z),
\ee
where $ P_j^{(1,1)}( \tilde z) $ are Jacobi polynomials and $ \tilde z = z/Z_0 $ is a dimensionless
variable.
 
Using Eq. (\ref{phase}) for the phase operator, $ F (\{ {\bf r}_i \}) $ becomes, 
\bearr
F( \{z_i \}) &=& \sum_{j=1}^{j_{{\rm max}}} \frac{2 a a_z^2}{Z_0 R_0^2} 
\frac{(j+2)(2j+3)}{(j+1) \sqrt{j(j+3)}} coth[\frac {\hbar \omega_j}{2k_BT}] 
\nonumber \\
& \times & [P_j^{(1,1)}(\tilde z_1)  +  P_j^{(1,1)}(\tilde z_2)
\nonumber \\ & - & ( P_j^{(1,1)}(\tilde z_3) + P_j^{(1,1)}(\tilde z_4))]^2.
\eearr
The upper cut-off limit ($j_{{\rm max}} $) on the summation can be obtained from the constraint:
$ \epsilon_j < \mu $. 
It is useful to define the space variables in the following way:
$ \tilde z_1 = (s+d)/2 = - \tilde z_2 $, and 
$\tilde z_3 = (s-d)/2 = - \tilde z_4$.
Here, $d$ is the displacement between  the two interfering condensate copies and $s$
is the relative distance between the positions in the interference pattern at which the
particle densities are evaluated. This new choice of variables  has been realized in the
experiment \cite{cacci}.

In Fig.1 and Fig.2, we plot the axial two-particle correlation function vs the relative separation at 
$T=0 $ and $ T=0.6 T_c $, respectively, for three different choices of the aspect ratio. 

\begin{figure}[h]
\epsfxsize 9cm
\centerline {\epsfbox{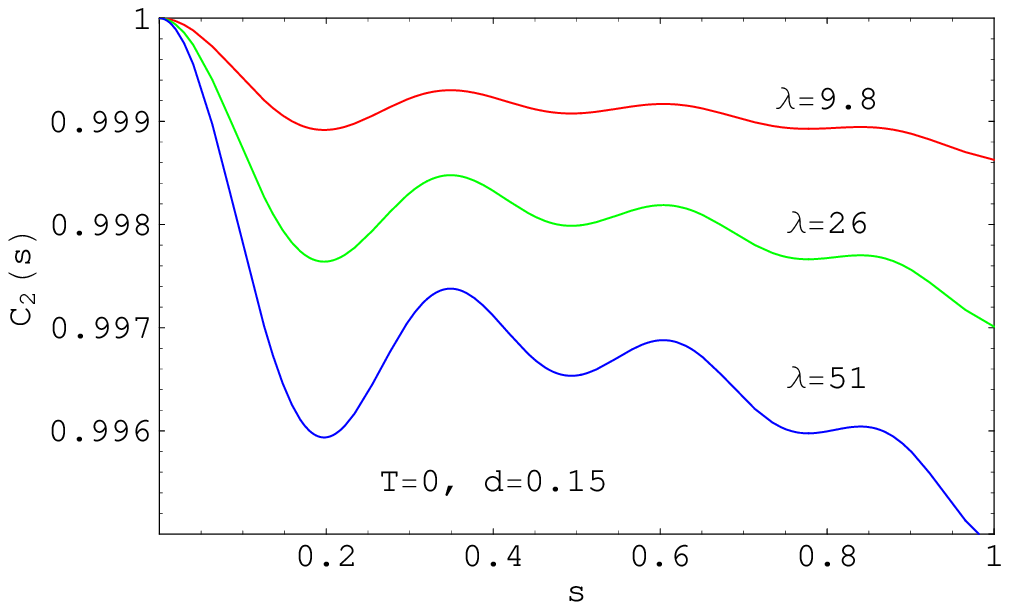}}
\vspace{0.2 cm}
\caption{(Color online) Plots of the axial two-particle correlation functions vs the relative separation.}
\end{figure}

\begin{figure}[h]
\epsfxsize 9cm
\centerline {\epsfbox{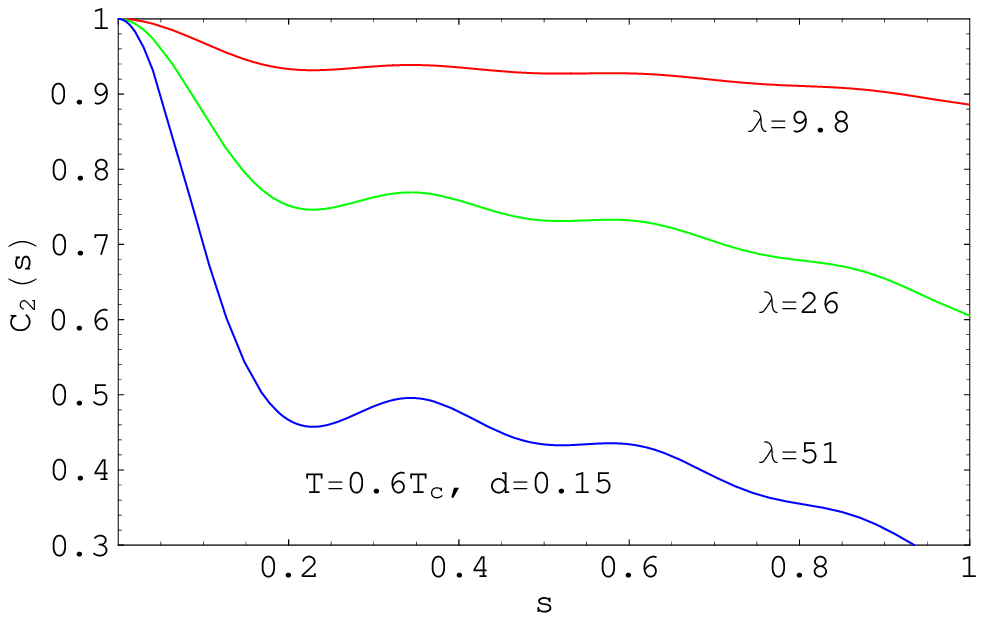}}
\vspace{0.2 cm}
\caption{(Color online) Plots of the axial two-particle correlation functions vs the relative separation.}
\end{figure}

The axial two-particle correlation function at zero temperature 
has oscillatory behavior with very small amplitude.
It shows little evidence of the stripes formation even at
zero temperature. Nevertheless, we do not claim that stripes are present 
even at $T=0$ because it is almost one for a large separation.
The minima of the $C_2(s) $ at various relative separations
are decreasing with the increasing of the aspect ratio $\lambda $. It implies 
that vacuum phase fluctuation increases with the aspect ratio.

Fig.2  shows the oscillatory behavior with large amplitudes compared to the zero temperature case.
These prominent deep valleys implies the presence of stripes in the axial density distribution due
to the strong thermally excited phase fluctuations and the large aspect ratio $\lambda $.
The minima of the $C_2(d,s) $ at various relative separations are decreasing with
the increasing of the temperature and the aspect ratio $\lambda $.

In the actual experiment \cite{exp}, the stripes formation is observed after the ballistic expansion of the
atomic cloud.  However, it does not rule out the possibility of the stripes formation in static BEC {\em
i.e.}, without ballistic expansion. 
According to the Ref. \cite{exp}, the observation of the stripes formation after the ballistic expansion of 
the cloud is due to the rapidly decreasing of the mean-field interaction and the
axial velocity fields are then converted into the density distribution.
In our studies we find that the stripes formation in the axial density 
profile is already present even before switching off the trap. 
Therefore, the stripes formation are not necessarily due to the axial velocity fields which are 
converted into density modulations during the ballistic expansion, as stated in Ref. \cite{exp}. 
The width of the stripes are too small to be observed directly in-trap due to the 
lack of the experimental resolution, but it is seen after 
expansion of the cloud since the stripes in the static cloud is enlarged during the 
ballistic expansion of the cloud.

{\em Quasi-1D Bose gas in the mean-field regime}:
Now we consider quasi-1D Bose gas in the mean-field regime as described in Ref. \cite{gorlitz}.
In particular, the system consists of $ N \sim 10^4 $ atoms of $^{23}$ Na in the trap 
with axial trapping frequency $ \omega_z = 2 \pi \times 3.5 $ Hz,
and radial trapping frequency $ \omega_0 = 2 \pi \times 360 $ Hz.
The low-energy excitation spectrum is given by $ \omega_j = \omega_z \sqrt{j(j+1)/2} $ 
and the corresponding normalized eigenfunctions are $ \psi_j (z) = \sqrt{\frac{2j+1}{2 Z_0}} P_j(z/Z_0) $,
\cite{ho} where $ P_j(z/Z_0) $ is the Legendre polynomial in $z$ and 
$Z_0 = a_z(2.998 Naa_z/a_0^2)^{1/3} $ is the Thomas-Fermi 
half length.

Using Eq. (\ref{phase}) for the phase operator, $ F (\{ {\bf r}_i \}) $ for quasi-1D system becomes,
\bearr 
F(\{z_i \}) &=& \sum_{j=1}^{j_{{\rm max}}} \frac{ a a_z}{a_0^2}  \frac{a_z}{Z_0}
\frac{(2j+1)}{ \sqrt{2j(j+1)}} coth[\frac{\hbar \omega_j}{2k_BT}] 
\nonumber \\
& \times & [P_j(\tilde z_1)  +  P_j(\tilde z_2) - P_j(\tilde z_3) - P_j(\tilde z_4)]^2.
\eearr
Using the expression for $ F (\{z_i \}) $, we plot two-point correlation functions vs the relative 
distance at $ T=0 $ and $ T=0.6 T_c $ in Fig.3 and Fig.4, respectively.

Similar to the 3D BEC in very elongated traps, the vacuum phase fluctuations are not able to
produce the stripes in quasi-1D Bose gas in the mean-field regime, However, at large temperature,
the two-point correlation functions have prominent oscillatory behavior with large amplitudes, compared
to the zero temperature case. Therefore, the stripes are present in the quasi-1D Bose gas in the
mean-field regime at large temperature.

\begin{figure}[h]
\epsfxsize 9cm
\centerline {\epsfbox{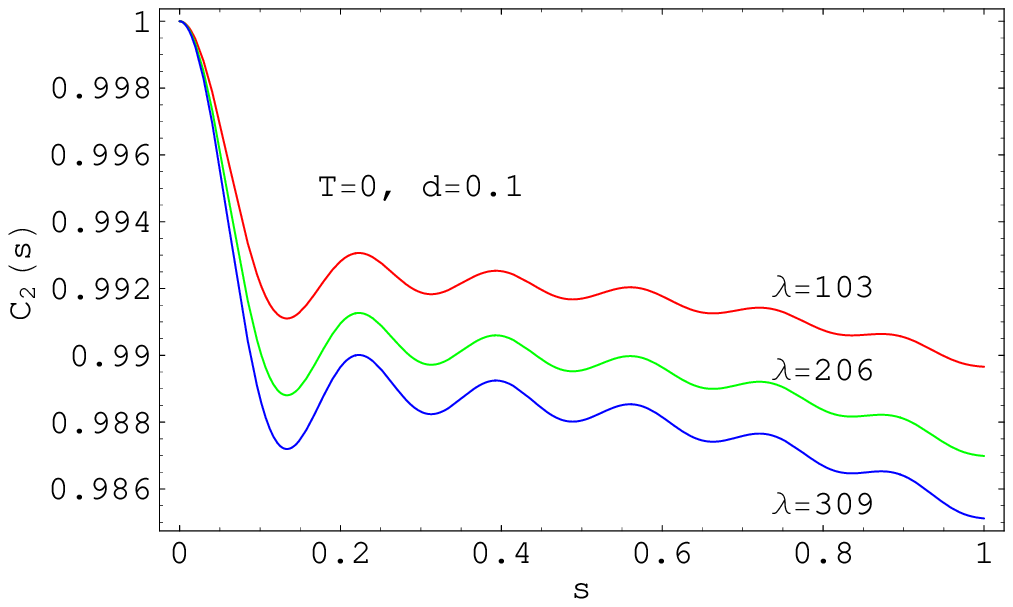}}
\vspace{0.2 cm}
\caption{(Color online) Plots of two-point correlation functions vs the relative distance for various trapping 
aspect 
ratio.}
\end{figure}

\begin{figure}[h]
\epsfxsize 9cm
\centerline {\epsfbox{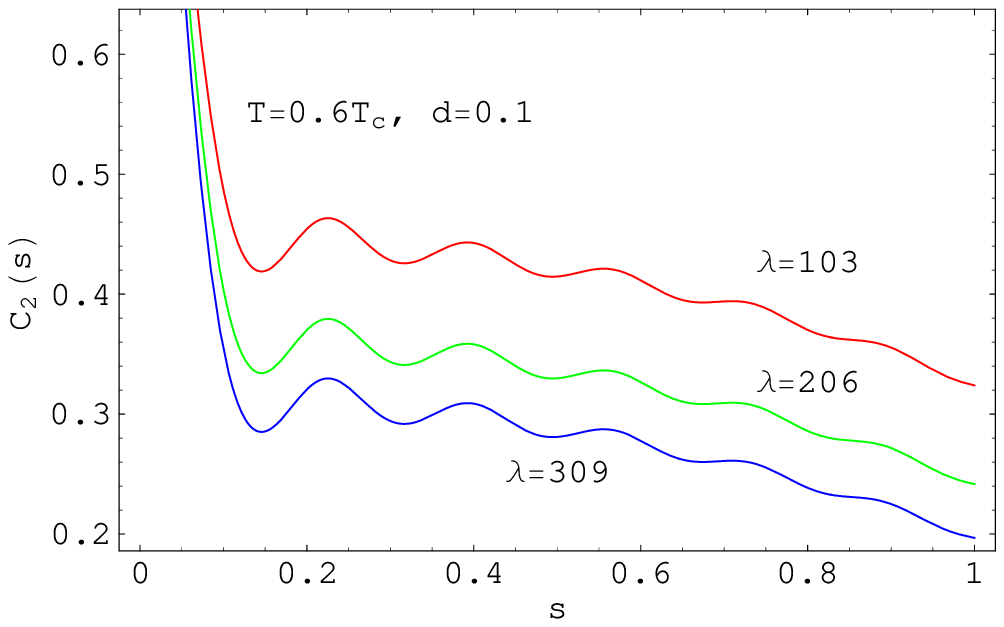}}
\vspace{0.2 cm}
\caption{(Color online) Plots of two-point correlation functions vs the relative distance for various trapping 
aspect ratio.}
\end{figure}

{\em Quasi-1D Bose gas in the hard-core bosons regime}:
Let us consider quasi-1D Bose gas in the hard-core bosons regime.
The low-energy excitation spectrum is given by $ \epsilon_j = j \hbar \omega_z $ and the
corresponding normalized eigenfunctions are 
$ \psi_j (z) = \sqrt{\frac{2}{\pi Z_0}} T_j(z/Z_0) $, where $ T_j(z/Z_0) $ is the first-order Chebyshev 
polynomial in $z$.
Here, $ Z_0 = \sqrt{2N} a_z $ is the
Thomas-Fermi half-length of the hard-core bosons system with the chemical potential 
$ \mu = N \hbar \omega_z $ \cite{tkg1}.  

Using Eq. (\ref{phase}) for the phase operator, $ F (\{ \bf r_i \}) $ for quasi-1D system in the hard-core 
bosons regime at $T=0 $ becomes
\be
F [z_i] = \sum_{j=1}^{N} \frac{1}{j} [ T_j(\tilde z_1) +  T_j(\tilde z_2)
- T_j(\tilde z_3) - T_j(\tilde z_4)]^2. 
\ee

Fig. 5 shows the two-point correlation function of the quasi-1D Bose gas in the hard-core regime 
at $T=0$.
\begin{figure}[h]
\epsfxsize 9cm
\centerline {\epsfbox{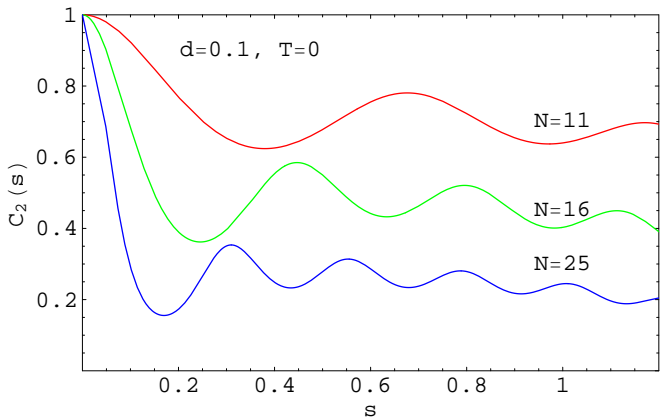}}
\vspace{0.2 cm}
\caption{(Color online) Plots of two-point correlation functions vs the relative distance for various number of 
hard-core atoms.}
\end{figure}

Note that Fig.5 is valid only when $ s > 1/2N $ due to the hydrodynamic approximation. This
hydrodynamic approximation is failed to describe the short-range correlations due to the strong
interactions when $ s < 1/2N $.
Fig. 5 shows that the two-point correlation functions have the oscillatory behavior with 
large amplitude even at zero temperature.
This implies that stripes are already present in the hard-core bosons regime even at zero temperature 
due to the strong quantum phase fluctuations. Note that the amplitude of the oscillations in the
zero temperature  two-point correlation functions is very very
small in the cigar-shaped 3D BEC as well as in the quasi-1D Bose gas in the mean field regime, compared to the
quasi-1D Bose gas in the hard-core regime.
The two-point correlation functions with large amplitude of the oscillations implies the fermionic like
behavior in the hard-core Bose gas at zero temperature.

{\em Detection}:
The realization of the stripes formation is possible by measuring the axial two-particle correlation
function. 
The Bragg interferometric method presented in Ref. \cite{cacci} to measure the two-particle correlation
function is
analogous to the original HBT experiment \cite{hbt}.
Therefore, the prominent valleys in the axial two-particle correlation function should be observe by using the Bragg
interferometric method as described in Ref. \cite{cacci} with a proper choices of the wave vector $k$ and 
the time interval ($\Delta t $) between the two Bragg pulses such that the distances 
($d = 2 \hbar k \Delta t/m $) between two auto correlated copies should be $ d \sim 0.1-0.2 $.
We have checked that for other choices of $d$, $ d \geq 0.2 $, 
the oscillatory behavior in the two-point correlation functions washes out.
This is due to the fact that the two interfering matter waves are not HBT-correlated when $ d \geq 0.2 $. 
Here we mean HBT-correlated in a sense that the matter waves will produce an HBT effect at the output
ports, as opposed to the different question of the correlations in the parent condensate \cite{baym}. In fact,
the two matter waves are HBT-correlated only if they are interfering within a coherence time ($\tau $),
the characteristic time scale in the experiment.
The relative distance $d$ between two interfering BECs at the output 
ports of the interferometer can be controlled by varying the time interval ($\Delta t $).
When $ \Delta t$ is small such that $d \sim 0.1-0.2$,  
we expect the correlations to be maximal and shows the HBT 
effect.
For large $ \Delta t > \tau $, the distance between two interfering BECs is large so that it becomes 
HBT-uncorrelated and then it does not produce valleys in the interference patterns. 
The time interval, $\Delta t $, before colliding two copies should be less than the coherence time $\tau $ 
which is 
inversely proportional to the temperature of the condensate {\em i.e.} $ \Delta t < \tau \sim \hbar /k_B T $,
where $T$ is the temperature of the parent condensate \cite{baym}.

In conclusion, we have 
understand the stripes in the axial density profile of a strong
phase fluctuating  3D cigar-shaped BECs in terms of the two-point correlation function. 
We have also predicted that
the stripes are present in the quasi-1D Bose gas in the mean-field regime at large temperature 
as well as in the 1D hard-core bosonic systems even at zero temperature due to to 
the strong quantum phase fluctuations.
We have also pointed out that one can realize the stripes 
by a proper choices of the parameters in the Bragg interferometer experiment which is similar to the
HBT setup. This experiment would provide the direct realization of the stripes formation, which is
of fundamental importance in the phase fluctuating BECs.

\end{multicols}

\begin{references}
\bibitem{rmp}
M. J. Anderson {\em et al.}, Science {\bf 269}, 198 (1995);
K. B. Davis {\em et al.}, Phys. Rev. Lett. {\bf 75}, 3969 (1995);
C. C. Bradley {\em et al.}, Phys. Rev. Lett. {\bf 78}, 985 (1997).

\bibitem{coh}
J. Stenger {\em et al.}, Phys. Rev. Lett. {\bf 82}, 4569 (1999);
E. W. Hagley {\em et al.}, Phys. Rev. Lett. {\bf 83}, 3112 (1999).

\bibitem{petrov}
D. S. Petrov {\em et al.}, 
Phys. Rev. Lett. {\bf 84}, 3745 (2000).

\bibitem{gorlitz}
A. Gorlitz {\em et al.}, Phys. Rev. Lett. {\bf 87}, 130402 (2001).

\bibitem{pet1}
D. S. Petrov  {\em et al.}, Phys. Rev. Lett. {\bf 85}, 3745 (2000).

\bibitem{tkg}
T. K. Ghosh, cond-mat/0402079

\bibitem{lieb}
M. Girardeau, J. Math. Phys. {\bf1}, 516 (1960);
E. H. Lieb and W. Liniger, Phys. Rev. {\bf 130}, 1605 (1963).

\bibitem{bloch}
B. Paredes, {\em et al.}, Nature {\bf 429}, 277 (2004).

\bibitem{exp}
S. Dettmer {\em et al.}, Phys. Rev. Lett. {\bf 87}, 160406 (2001).

\bibitem{hbt}
R. Hanbury-Brown and R. Q. Twiss, Nature(London) {\bf 177}, 27 (1956);
R. Hanbury-Brown and R. Q. Twiss, Nature(London) {\bf 178}, 1046 (1956).

\bibitem{baym}
G. Baym, Acta Phys. Polon. B {\bf 29}, 1839 (1998).

\bibitem{cacci}
L. Cacciapuoti {\em et al.}, Phys. Rev. A {\bf 68}, 053612 (2003).

\bibitem{strin}
M. Fliesser {\em et al.},  Phys. Rev. A {\bf 56}, R2533 (1997);
S. Stringari, Phys. Rev. A {\bf 58}, 2385 (1998).

\bibitem{ho}
T.-L. Ho and M. Ma, J. of Low Temp. Phys. {\bf 115}, 61 (1999).

\bibitem{tkg1}
T. K. Ghosh, cond-mat/0402080.


\end{references}
\end{document}